# The Rise and Rise of Citation Analysis

## Lokman I. Meho

School of Library and Information Science, Indiana University - 1320 East 10th St., LI 011; Bloomington, IN 47405; Tel: 812-855-2018; meho@indiana.edu

It is a sobering fact that some 90% of papers that have been published in academic journals are never cited. Indeed, as many as 50% of papers are never read by anyone other than their authors, referees and journal editors. We know this thanks to citation analysis, a branch of information science in which researchers study the way articles in a scholarly field are accessed and referenced by others (see box 1).

Citation analysis is, however, about much more than producing shock statistics. Along with peer review, it has over the past three decades been increasingly used to judge and quantify the importance of scientists and scientific research. Citation analysis is also the machinery behind journal "impact factors" – figures of merit that researchers take note of when deciding which journal to submit their work to so that it is read as widely as possible. Indeed, the output from citation studies is often the only way that non-specialists in governments and funding bodies – or even those in different scientific disciplines – can judge the importance of a piece of scientific research.

The Web has had a huge impact on citation-analysis research. Since the turn of the century, dozens of databases

> **BOX 1: Citation Analysis at a Glance**
>
> - Citation analysis, which involves counting how many times a paper or researcher is cited, assumes that influential scientists and important works are cited more often than others.
> - Although the ISI Science Citation Index has long been the most common tool for measuring citations in physics, other Web-based services are now challenging its dominance.
> - Each service produces slightly different results, revealing the importance of using several citation sources to judge the true impact of a scientist's work.
> - The Web is also leading to alternatives to the traditional "impact factor" of a journal or individual, including download counts and the *h*-index.

such as *Scopus* and *Google Scholar* have appeared, which allow the citation patterns of academic papers to be studied with unprecedented speed and ease. This could mark the beginning of the end for the 40-year monopoly of citation analysis held by the US-based firm Thomson Scientific, formerly known as the Institute for Scientific Information (ISI).

The ISI's citation databases have always been criticized by scientists on the basis that they index a limited number of journal titles, that they cover mainly English-language titles from North America and Western Europe, and that they do not cover citations from books and most conference proceedings. However, the myriad of Web-based sources now provides a much more comprehensive coverage of the world's literature, helping to usher in a new era of citation analysis based on multiple sources. Furthermore, the Web has led to several new citation measures and methods that were previously impractical, including article-download counts, link analysis, Google's PageRank, Web citations and the "$h$-index" recently developed by US physicist Jorge Hirsch.

**Out with the old**

Citation analysis essentially involves counting the number of times a scientific paper or scientist is cited, and it works on the assumption that influential scientists and important works will be cited more frequently than others. Many governments, funding agencies (in the US at least) and tenure and promotion committees use citation data to evaluate the quality of a researcher's work, partly because they prefer not to rely on peer review and publication output alone.

However, not everybody thinks citation analysis is the best way to judge the validity of a scientific claim. Critics point to basic citing errors such as "homographs", i.e., failing to separate citations to two unrelated scientists who happen to share the same last name and first initial.

Cronyism, whereby friends or colleagues reciprocally cite each other to mutually build their citation counts, is another drawback. Other problems include people deliberately citing themselves or journals they are involved with; ceremonial citations, in which an author cites an authority in the field without ever having consulted the relevant work itself; and negative citations pointing out incorrect results.

Proponents of citation analysis, on the other hand, claim that these problems are relatively insignificant. Most citations found in articles and books, they say, are useful – by paying homage to pioneers, identifying original publications, providing background reading and alerting readers to forthcoming works. Citations also substantiate claims, give credit to related works and provide leads to poorly disseminated, poorly indexed or uncited works. According to Michael Koenig in the Palmer School of Library and Information Science at Long Island University in the US, citations provide – despite their ambiguities – "an objective measure of what is variously termed productivity, significance, quality, utility, influence, effectiveness, or impact of scientists and their scholarly products."

The ISI citation databases – which include the *Arts and Humanities Citation Index* (A&HCI), *Science Citation Index* (SCI) and *Social Sciences Citation Index* (SSCI) – have for decades been used as a starting point and often as the only tools for conducting citation analyses. Since their original publication in the 1960s and 1970s these databases have grown dramatically in size and influence, and today contain about 40 million records from more than 8700 of the world's most prestigious research journals. The SCI, which was launched in 1964, quickly became popular with scientists and librarians, and is now one of the most important multidisciplinary databases in the world.

Young researchers might find it hard to comprehend, but until 1988 these indexes existed only in print form, although searching them online has been possible since the mid 1970s using third-party information-retrieval systems such as Dialog. In 1988 the ISI supplemented its indexes with CD-ROM editions, and in 1997 the databases finally migrated online with the launch of *Web of Science*. The move to an online interface, which can analyse thousands of records in a few seconds, has given the ISI's databases an even greater stranglehold in the field of citation analysis. But at the same time the Web has produced new publication venues and competitors that challenge the wisdom of continuing to use *Web of Science* exclusively.

Another problem with *Web of Science* is that it ignores the fact that scientists increasingly publish or "post" their papers online via open-access journals, personal homepages, e-print servers or in institutional repositories so that others can freely access the material. At the same time, researchers have started to search and download research materials via services such as *arXiv.org*, *Google Scholar* or publishers' websites, like Elsevier's *ScienceDirect*. Many of the millions of documents accessible via these services, which are published instantly to give the wider scientific community time to use and ultimately cite them, are not indexed by *Web of Science*. Moreover, an increasing number of Web-based services are enabling explicit citation searching (see box 2).

## Multiple citation sources

The rise in the use of Web-based databases and tools to access scientific literature has revealed how vital it is to use multiple citation sources to make accurate assessments of the impact and quality of scientists' work. Take the book *Quantum Computation and Quantum Information* by M Nielsen and I Chuang (2000 Cambridge University Press), for example. According to *Web of Science*, this book has been cited more than 2800 times. However, *Scopus* says it has been cited 3150 times, *Google Scholar* 4300 times, *Physical Review Online Archive* 150 times, *ScienceDirect* 375 times, the Institute of Physics *Journal Archive* 290 times, and *arXiv.org* 325 times. If only *Web of Science* is used, we would miss all of the citations found through *Google Book Search* and *arXiv.org* plus hundreds of the citations found through the other databases or tools.

A citation study carried out recently by the present author and Kiduk Yang at Indiana University in the US is one of many that have shown the need to use multiple citation sources. We compared results of citation coverage from *Web of Science*, *Scopus* and *Google Scholar* for a sample of 25 highly published researchers in our

---

**BOX 2: Web-Based Citation Analysis Tools**

The Web has given birth to more than 100 new databases or tools that allow citation searching. These fall into three categories. The first allows the user to search in the full-text field to determine whether certain items, authors or journals have been cited in a document. Examples of these databases or tools include the following:
- *arXiv* e-print server (arXiv.org)
- *CiteSeer* (citeseer.ist.psu.edu)
- *Google Book Search* (books.google.com)
- *Google Scholar* (scholar.google.com)
- Institute of Physics *Journal Archive* (journals.iop.org/archive)
- *Physical Review Online Archive* (prola.aps.org)
- Elsevier's *Scirus* (scirus.com)

Some of these tools, such as *CiteSeer* and *Google Scholar*, are based on autonomous citation indexing that allows automatic extraction and grouping of citations for online research documents. The second category of databases or tools allows the user to search in the cited references field to identify relevant citations. These tools first became available in the late 1990s when subject-specific databases began adding cited-references information to their own records. Examples include the following:
- NASA's *Astrophysics Data System Abstract Service* (adsdoc.harvard.edu)
- *MathSciNet* from the American Mathematical Society (ams.org/mathscinet)
- Elsevier's *ScienceDirect* (sciencedirect.com)
- *SciFinder Scholar* from the American Chemical Society (cas.org)
- *Scitation/SPIN* from the American Institute of Physics (scitation.aip.org)
- *SPIRES-HEP* at Stanford Linear Accelerator Center (slac.stanford.edu/ spires)

The third category is databases that work exactly like *Web of Science*. The main and perhaps only good example of this category is *Scopus* (scopus.com), which was launched in 2004 by Elsevier. Although it covers more refereed journals and conference proceedings than *Web of Science* (15,000 titles compared with 8,700) *Scopus* provides citation searching only from 1996 onwards, whereas *Web of Science* goes as far back as 1900.

field of information science and found that *Scopus* and *Google Scholar* increase the citation counts of scholars by an average of 35% and 160%, respectively. Perhaps more importantly, this increase varies considerably from one research area to another, with researchers working in computer-mediated communication and human–computer interaction having their number of citations more than doubled, while those specializing in bibliometrics and citation analysis had their number of citations increase by less than 25%.

Another major finding of our study is that the use of *Scopus* and *Google Scholar* has helped to establish a link between information-science research and cognitive science, computer science, education and engineering (as evidenced by the high number of citations from journal articles and conference papers in these fields). Such a finding about interdisciplinary trends in science could not have been uncovered by relying on *Web of Science* citations only. Indeed, multiple citation tools allow us to generate much more accurate maps or visualizations of scholarly communication networks in general, such as establishing links between authors, departments, disciplines, journals or countries that cite or influence each other.

While the emergence of comprehensive Web-based citation databases and tools – many of which have been around for only two years or so – has been received favourably by citation analysts, it has also made the job of searching and analysing citations more challenging. For instance, the new citation tools cover not only journal and conference papers, but also millions of unique items in various languages and forms, such as book chapters, dissertations, e-prints and research reports. This requires much more work than the relatively simple task of using *Web of Science* to compile and interpret citation searching and analyses based mainly on refereed journal articles.

For instance, we spent about 3000 hours collecting data from *Google Scholar* alone in order to carry out our recent study into the overlap and uniqueness between citation databases, compared with only 100 hours using *Web of Science* and 200 using *Scopus* for the same sample. Another consideration when performing citation analyses in the Web era is how to weigh up citations from journal versus non-journal sources and from refereed versus non-refereed sources. This is vital because citations from, say, the journal *Nature Physics* are of different quality and value than citations found in a Masters thesis that sits in a university repository.

## Quantifying your impact

The basic idea of using citation-based measures to assess the impact, importance or quality of a scientist's overall work is to show how often and where he or she is cited. The two best known measures are citation counts and "impact factor" – the number of citations received in the current year to articles published in the two preceding years divided by the total number of articles published in

**BOX 3: The Impact Factor**

When scientists seek research grants, file for tenure or promotion, or apply for staff or faculty positions, it has become customary to include both the impact factor scores of the journals in which their papers were published and the number of citations received by these articles. As high-impact journals usually attract high-quality contributions from top scientists and have a large readership, publishing in these journals is a top priority for scientists who want to increase their visibility, prestige and influence among their peers; it also improves their chances of getting lucrative job offers and research grants.

The impact factor of a journal in a particular year is the number of citations received in the current year to articles published in the two preceding years divided by the number of articles published in the same two years. For example, *Physical Review Letters* has a 2005 impact factor of 7.489, which means that on average each of its 2003 and 2004 articles was cited 7.489 times in 2005.

The journal impact factor was launched by the Institute for Scientific Information (ISI) in 1975 and has been published annually since then. Initially, the impact factor was made available in microfiche format only, then ISI migrated it to CD-ROM in the late 1980s, before finally making a searchable version of it available on the Web in 1997. Currently, the ISI provides impact-factor data for over 5,900 journals in science and technology and 1,700 journals in the social sciences through the publication *Journal Citation Reports*. The impact factors always lag one year behind and are published each summer.

the same two years (see box 3). However, the fact that almost all scientific papers now exist online opens the door to other, possibly more accurate, citation-based measures.

The impact factor has several weaknesses. First, its scores can be significantly influenced by a few highly cited articles and/or too many uncited or low-cited articles. Second, authors and journals that frequently publish review articles tend to have their citation counts and impact exaggerated because these types of articles are usually highly cited. Third, citation counting and impact factors do not take into account articles that were used but did not get cited. Finally, the two-year "citation window" in the impact-factor formula fails to capture the "long-term value" or the real impact of many journals.

The Web has enabled a number of alternative citation- based measures to be devised to get around some of the limitations of the citation-counting and impact-factor methods. Of these, the most important are "download counts", which became feasible only because of the migration to online publication, and the "*h*-index", which exploits the rise of Web-based citation databases.

Using a download rather than citation count means that the impact of an article or a journal can be measured in real time, rather than having to wait several years after it has been published. According to Tim Brody and Steven Harnad in the School of Electronics and Computer Science at the University of Southampton in the UK, there is a strong, positive correlation between download counts and both citation counts and impact factors, although the degree of correlation varies from one research field to another. As downloads are instantly recorded and counted, Brody and Harnad suggest that the measure can be particularly useful for providing an early estimate of the probable citation impact of articles.

The *h*-index, meanwhile, was developed in 2005 by Jorge Hirsch, a condensed-matter physicist at the University of California in San Diego, to quantify the impact and quality of individual scientists' research output (see box 4). The measure is simple: a scientist with an *h*-index of, say, 40 has published 40 articles that have each attracted at least 40 citations (papers with fewer than 40 citations therefore do not count). Hirsch estimates that after 20 years a "successful scientist" will have an *h*-index of 20, an "outstanding scientist" an *h*-index of 40, and a "truly unique" individual an *h*-index of 60. However, he points out that values of *h* will vary between different fields.

The *h*-index – which ranks Ed Witten of Princeton University top with an *h* value of 107 – immediately became popular among researchers because it captures the fact that scientists with very few high-impact articles or, alternatively, many low-impact articles will have a low *h*-index. The measure therefore helps distinguish between a "one-hit wonder" and an enduring performer who has numerous high-impact articles and hence a high *h*-index.

> **BOX 4: How high is your *h*-index?**
>
> In 2005 the US condensed-matter physicist Jorge Hirsch devised a simple metric with which to quantify the scientific output of an individual: a scientist has an *h*-index of 10, say, if he or she has published 10 papers that have received at least 10 citations each. To compute your own *h*-index you must first identify all the relevant records in a citation database (see box on page XX) in which you are an author and then automatically (or manually if you use *Google Scholar*) sort the records by the number of times they have been cited, with the most cited listed first. To calculate *h* all you then have to do is count down until the number of records equals or is no longer greater than the number of times cited. Although originally meant for quantifying the impact and quality of individual scientists' research output, the *h*-index has been successfully applied to journals, research projects and entire research groups.

Moreover, a flurry of empirical studies conducted by librarians and others shows that the *h*-index correlates positively with citation counts, impact factors, publication counts and peer evaluation of research impact and quality. Finally, the *h*-index is very easy and quick to compute using databases or tools such as *Web of Science*, *Scopus* or *Google Scholar*. Indeed, almost an entire issue of the journal *Scientometrics* was recently devoted to the *h*-index, and the measure is now automatically calculated in the "citation report" function of *Web of Science*.

Like all citation-based measures, however, the *h*-index must be used with caution. This is because the index ignores, for example, why an item was cited in the first place; so negative citations to incorrect work are counted. Moreover, it is insensitive to highly cited works and disregards total citation counts. These last drawbacks have led to the development of two alternative indexes. The editor of *Science Focus* Jin Bihui has devised an "*a*-index", which is defined as the average number of citations received by works in the number of *h*-index publications, while Leo Egghe from the Universiteit Hasselt in Belgium has devised a "*g*-index", which is defined as the highest number, *g*, of papers that together received $g^2$ or more citations. (A researcher with a *g*-index of, say, 10 has published 10 papers that together have been cited at least 100 times.)

Indeed, Hirsch seems to have encouraged other physicists to develop their own productivity measures, with terms such as the "*h–b* index" and "creativity index, *Ca*" having appeared in preprints on the *arXiv* server in recent months. The *h–b* index was devised by Michael Banks from the Max-Planck Institute for Solid-State Physics in Stuttgart, Germany, to judge the impact of a particular field, whereas the creativity index was developed by José Soler of the Universidad Autonoma de Madrid in Spain to create and transmit scientific knowledge based on the network of citations among research articles. The creativity index concluded that Princeton University Nobel laureate Philip Anderson is the most creative physicist in the world (*Physics World*, September 2006 p9).

**Harnessing the Potential**

The citation databases, tools and citation methods mentioned here are just a few of many new and powerful indicators of research output that have become possible with the Web. Indeed, a search for articles in *Web of Science* reveals that the number of citation-based research-evaluation studies has been growing steadily over the years. Meanwhile, the exponential increase in the number of databases and tools that allow citation searching shows just how widespread the use and popularity of citation analysis has become. Funding agencies, as well as hiring and promotion committees, are increasingly relying on these methods to evaluate research to supplement other quality indicators such as peer review and publication output.

The Web has brought many changes and challenges to the field of citation analysis. Researchers and administrators who want to evaluate research impact and quality accurately will from now on have to use not only multiple sources – *Web of Science* and *Scopus* being the main two, but also *Google Scholar*, *arXiv.org* and others – but also different methods (e.g. citation counts as well as the *h*-index, and so on) to corroborate their findings. Relying exclusively on *Web of Science* and a single citation measure will, in many cases, no longer be an option for making accurate impact assessments.

Scientists now need to make it their job to disseminate their work on as many platforms and in as many different ways as possible, such as publishing in open access and high-impact journals, and posting their work in institutional repositories, personal homepages and e-print servers, if they want their peers to be aware of, use and ultimately cite their work. Publishing a journal article is now only the first step in disseminating or communicating one's work; the Web provides a multitude of methods and tools to publicize its scholarly worth.

## More About Citation Analysis: